\documentclass[onecolumn,showpacs,10pt]{revtex4}

\usepackage{graphicx}
\usepackage{dcolumn}
\usepackage{bm}

\input epsf

\begin{document}

\title{\Large Thermodynamics in Higher Dimensional Vaidya Space-Time}

\author{\bf Ujjal Debnath\footnote{ujjaldebnath@yahoo.com ,
ujjal@iucaa.ernet.in}}

\affiliation{Department of Mathematics, Bengal Engineering and
Science University, Shibpur, Howrah-711 103, India.\\}

\date{\today}

\begin{abstract}
In this work, we have considered the Vaidya spacetime in null
radiating fluid with perfect fluid in higher dimension and have
found the solution for barotropic fluid. We have shown that the
Einstein's field equations can be obtained from Unified first law
i.e., field equations and unified first law are equivalent. The
first law of thermodynamics has also been constructed by Unified
first law. From this, the variation of entropy function has been
derived on the horizon. The variation of entropy function inside
the horizon has been derived using Gibb's law of thermodynamics.
So the total variation of entropy function has been constructed at
apparent and event horizons both. If we do not assume the first
law, then the entropy on the both horizons can be considered by
area law and the variation of total entropy has been found at both
the horizons. Also the validity of generalized second law (GSL) of
thermodynamics has been examined at both apparent and event
horizons by using the first law and the area law separately. When
we use first law of thermodynamics and Bekenstein-Hawking area law
of thermodynamics, the GSL for apparent horizon in any dimensions
are satisfied, but the GSL for event horizon can not be satisfied
in any dimensions.
\end{abstract}


\maketitle

\section{\normalsize\bf{Introduction}}

The connection between gravity and thermodynamics was discovered
by Jacobson \cite{Jacob} by deriving the Einstein equation
together with the first law of thermodynamics where the entropy is
proportional to the horizon area. The horizon area of black hole
is associated with its entropy and the surface gravity which is
related with its temperature in black hole thermodynamics
\cite{Beken,Haw,Bard}. Frolov et al \cite{Frol} calculated the
energy flux of a background slow-roll scalar field through the
quasi-de Sitter apparent horizon and used the first law of
thermodynamics $-dE = TdS$, where $dE$ is the amount of the energy
flow through the apparent horizon. Using the Hawking temperature
$T_{A}=\frac{1}{2\pi R_{A}}$ and Bekenstein entropy
$S_{A}=\frac{\pi R_{A}^{2}}{G}$ ($R_{A}$ is the radius of apparent
horizon) at the apparent horizon ($G$ is the Newton's constant),
the first law of thermodynamics (on the apparent horizon) is shown
to be equivalent to Friedmann equations \cite{Cai1} and from this,
the generalized second law (GSL) of thermodynamics is obeyed at
the horizon. Gibbons and Hawking \cite{Gibb} first investigated
the thermodynamics in de Sitter space–time. Verlinde \cite{Verl}
found that the Friedmann equation in a radiation dominated FRW
universe can be written in an analogous form of the Cardy-Verlinde
formula. Padmanabhan \cite{Pad} was formulated the first law of
thermodynamics on the horizon, starting from Einstein
equations for a general static spherically symmetric space time.\\

In a spatially flat de Sitter space–time, the event horizon and
the apparent horizon of the Universe coincide and there is only
one cosmological horizon for this space time. When the apparent
horizon and the event horizon of the Universe are different, it
was found that the first law and generalized second law (GSL) of
thermodynamics hold on the apparent horizon, while they break down
if one considers the event horizon \cite{Wang1}. Considering FRW
model of the universe, most studies deal with validity of the
generalized second law of thermodynamics starting from the first
law when universe is bounded by the apparent horizon
\cite{Bous,Cai2,Akb}. In the reference \cite{Izq}, a Chaplygin gas
dominated expansion was considered and the GSL was investigated
taking into account the existence of the observer's event horizon
in accelerated epanding universe and from this, it was concluded
that for the initial stage of Chaplygin gas dominated expansion,
the GSL of gravitational thermodynamics is satisfied.\\

Most of the cosmological thermodynamics have been studied in open,
closed and flat FRW models of the universe. Recently, the
generalized second law of thermodynamics (GSL) have been studied
for spherically symmetric LTB model \cite{Chak} in four and higher
dimensional quasi-spherical Szekeres' model \cite{Deb1}. Unified
first law and thermodynamics of dynamical black hole in
$n$-dimensional Vaidya spacetime have been discussed in
\cite{Rong}. Generalized Vaidya spacetime in Lovelock gravity and
thermodynamics on the apparent horizon have also been studied
\cite{Cai3}. In 1996, Husain \cite{Hus} gave non-static
spherically symmetric solutions of the Einstein equations for a
null fluid source with pressure $p$ and density $\rho$ related by
the equation of state $p=k\rho$. Wang et al \cite{Wang2} has
generalized the Vaidya solution which include most of the known
solutions to the Einstein equation such as anti-de-Sitter charged
Vaidya solution. Husain solution has been used to study the
formation of a black hole with short hair \cite{Brown} and can be
considered as a generalization of Vaidya solution \cite{Wang2}.\\

In this work, we briefly describe the generalization of Vaidya
solution (i.e., Husain solution) in $(n+2)$-dimensional
spherically symmetric space-time \cite{Deb2} for a null fluid
source with barotropic fluid. Using unified first law, the first
law of thermodynamics has been derived in higher dimensional
Vaidya spacetime model. The expressions of apparent horizon and
event horizon radius have been obtained. The surface gravity and
temperature on the apparent and event horizons have been
calculated. The validity of GSL of thermodynamics has been
examined using first law and area law of thermodynamics separately
in presence of barotropic fluid with Vaidya null radiation.\\

\section{ \normalsize\bf{Brief Review of} \large ($n$+2)\normalsize\bf{-Dimensional Vaidya
Space-Time}}

The spherically symmetric inhomogeneous metric in
$(n+2)$-dimensional space-time can be taken as \cite{Deb2}

\begin{equation}
ds^{2}=-\left(1-\frac{m(v,r)}{r^{n-1}}\right)dv^{2}+2dvdr+r^{2}d\Omega_{n}^{2}
\end{equation}

where $r$ is the radial co-ordinate ($0<r<\infty$), $v$ is the
null co-ordinate ($-\infty\le v\le \infty$) which stands for
advanced Eddington time co-ordinate, $m(v,r)$ gives the
gravitational mass inside the sphere of radius $ r$ and
$d\Omega_{n}^{2}$ is the line element on a unit $n$-sphere. For
the matter field, we now consider two non-interacting components
of energy momentum tensors namely the Vaidya null radiation and a
perfect fluid having form

\begin{equation}
T_{\mu\nu}=T_{\mu\nu}^{(n)}+T_{\mu\nu}^{(m)}
\end{equation}
where
\begin{equation}
T_{\mu\nu}^{(n)}=\sigma l_{\mu}l_{\nu}
\end{equation}

is the component of the matter field which moves along null
hypersurface and

\begin{equation}
T_{\mu\nu}^{(m)}=(\rho+p)(l_{\mu}\eta_{\nu}+l_{\nu}\eta_{\mu})+pg_{\mu\nu}
\end{equation}

represents the energy-momentum tensor of matter. Here $\rho$ and
$p$ are the energy density and thermodynamic pressure while
$\sigma$ is the energy density corresponding to Vaidya null
radiation. In the comoving co-ordinates
($v,r,\theta_{1},\theta_{2},...,\theta_{n}$), the two eigen
vectors of energy-momentum tensor namely $l_{\mu}$ and
$\eta_{\nu}$ are linearly independent future pointing null vectors
having components

\begin{equation}
l_{\mu}=(1,0,0,...,0)~ \text{and}~  \eta_{\mu}=\left(\frac{1}{2}\left(1-\frac{m}{r^{n-1}}\right),-1,0,...,0 \right)
\end{equation}

and they satisfy the following relations

\begin{equation}
l_{\lambda}l^{\lambda}=\eta_{\lambda}\eta^{\lambda}=0,~ l_{\lambda}\eta^{\lambda}=-1
\end{equation}

The non-vanishing components of the Einstein's field equations
(choosing $8\pi G=c=1$)

\begin{equation}
G_{\mu\nu}=T_{\mu\nu}
\end{equation}

for the metric (1) with matter field having stress-energy tensor
given by (2) are obtained as

\begin{equation}
\rho=\frac{nm'}{2 r^{n}},~~p=-\frac{m''}{2 r^{n-1}}~~\text{and}~ \sigma=\frac{n\dot{m}}{2 r^{n}}
\end{equation}

where an dot and dash stand for partial derivatives with respect
to $v$ and $r$ respectively. Now we assume the matter fluid
satisfies the barotropic equation of state

\begin{equation}
p=k\rho,~~~(k<1,~\text{a~constant})
\end{equation}

From (8) and (9), we obtain the explicit solution for the
gravitational mass function $m(v,r)$ as \cite{Deb2}

\begin{equation}
m(v,r)=\left\{
\begin{array}{lll}
f(v)-\frac{g(v)}{(nk-1)r^{nk-1}}~,~~~nk\ne 1\\
\\
f(v)+g(v)\log~r~,~~~nk=1
\end{array}\right.
\end{equation}

where the integration functions $f(v)$ and $g(v)$ are arbitrary
functions of $v$ alone. This generalized Vaidya solution is also
known as {\it Husain solution} in $(n+2)$-dimensions. Since
$\dot{m}=\dot{f}(v)+\dot{g}(v)~\log~r$~ becomes negative (i.e.,
$\sigma<0$) for very small $r$, so the energy conditions are not
always satisfied for all $r$ for the solution with $k=1/n$. Hence
we shall not consider this solution in future, because this is
physically unrealistic. Since $f(v)$ and $g(v)$ are arbitrary
functions of $v$, so without any loss of generality we may assume
[22] $f(v)=f_{0}v^{n-1}$ and $g(v)=g_{0}v^{n(k+1)-2}$, where
$f_{0}$
and $g_{0}$ are chosen to be positive.\\

\section{\normalsize\bf{Study of Thermodynamics in Vaidya Space-Time}}

In this section, we shall discuss the unified first law for higher
dimensional Vaidya space-time. In this model, the first law of
thermodynamics can be generated from unified first law. The
validity of generalized second law of thermodynamics will be
examined for apparent and event horizons using first law and using
area law in the subsequent sections. Now we consider the metric
(1) in the double null form \cite{Cai3,Cai4}

\begin{equation}
ds^{2} =h_{ab}dx^{a}dx^{b}+r^{2}d\Omega_{n}^{2}~~,~~a,b=0,1
\end{equation}

where
$h_{ab}=\left(-\left(1-\frac{m(v,r)}{r^{n-1}}\right),1,1,0\right)$.
The unified first law is defined by \cite{Cai3} the following form

\begin{equation}
dE={\cal A}\Psi+WdV
\end{equation}

where the surface area ${\cal A}$ is given by

\begin{equation}
{\cal A}=(n+1)\Omega_{n+1} r^{n}
\end{equation}

and the volume $V$ is defined by \cite{Cai1}

\begin{equation}
V=\Omega_{n+1} r^{n+1}~~ \text{where} ~~
\Omega_{n+1}=\frac{\pi^{\frac{n+1}{2}}}{\Gamma(\frac{n+3}{2})}
\end{equation}

From (2), we get the components of energy-momentum tensor as

\begin{equation}
T_{00}=\sigma+\rho\left(1-\frac{m(v,r)}{r^{n-1}}\right),~T_{01}=T_{10}=-\rho,~T_{11}=0
\end{equation}

The work density function $W$ is given by

\begin{equation}
W=-\frac{1}{2}h^{ab}T_{ab}=\rho
\end{equation}

The energy-supply vector is given by

\begin{equation}
\Psi_{a}=h^{bc}T_{ac}\partial_{b} (r)+W\partial_{a} (r)=(\sigma,0)
\end{equation}

So we have the energy flux or momentum density in the Vaidya
spacetime model as
\begin{equation}
\Psi=\Psi_{a}dx^{a}=\sigma dv
\end{equation}

The Misner-Sharp energy $E$ inside the Vaidya space-time surface
is given by \cite{Ren}

\begin{equation}
E=\frac{n(n+1)}{2}\Omega_{n+1}
r^{n-1}[1-h^{ab}\partial_{a}(r)\partial_{b}(r)]=\frac{n(n+1)}{2}\Omega_{n+1}
m(v,r)
\end{equation}

Now using (13), (14), (16) and (18), we get

\begin{equation}
{\cal A}\Psi+WdV=(n+1)\Omega_{n+1}r^{n}(\sigma dv+\rho dr)
\end{equation}

Taking the total differential of (19), we have

\begin{equation}
dE=\frac{n(n+1)}{2}\Omega_{n+1}(\dot{m}dv+m'dr)
\end{equation}

Using (20), (21) and the unified first law (12), comparing the
coefficients of $dv$ and $dr$, we obtain

\begin{equation}
\rho=\frac{nm'}{2 r^{n}}~~\text{and}~ \sigma=\frac{n\dot{m}}{2
r^{n}}
\end{equation}

which are the two field equations given in (8) for Vaidya
space-time. But it is not possible to find the field equation
$p=-\frac{m''}{2 r^{n-1}}$ directly from the unified first law. If
conservation of energy is considered then using (22) we shall get
this field equation. So we may conclude the unified first law and
the Einstein's field equations of $(n+2)$ dimensional Vaidya space-time are equivalent.\\

Now the Gibb's law of thermodynamics states that \cite{Wang1}

\begin{equation}
T_{h}dS_{I}=pdV+d(E_{I})
\end{equation}

where, $S_{I},~p,~V$ and $E_{I}$ are respectively entropy,
pressure, volume and internal energy within the horizon of the
Vaidya spacetime. Here the expression for internal energy can be
written as $E_{I}=\rho V$. Since the null radiation moves along
null hypersurface, so $E_{I}$ does not depend on null radiation
density $\sigma$. Here we consider the equilibrium thermodynamics,
so the temperature inside
the horizon $=$ temperature on the horizon $= T_{h}$.\\

Using (8) and (14), the above expression (23) can be simplified to
the form

\begin{equation}
\dot{S}_{I}=\frac{\Omega_{n+1}}{2T_{h}}\left[\{nm'-(n+1)m''r_{h}\}\dot{r}_{h}
+n\dot{m}'r_{h}\right]
\end{equation}

In the following two sections, we shall examine the validity of
generalized second law (GSL) of thermodynamics of the universe
bounded by apparent and event horizons using first law and area
law of thermodynamics separately for higher dimensional Vaidya
spacetime in presence of barotropic fluid with null radiation.\\

\section{\normalsize\bf{GSL using first law}}

We know that heat is one of the form of energy. Therefore, the
heat flow $\delta Q$ through the horizon is just the amount of
energy crossing it during the time interval $dv$. That is, $\delta
Q=-dE$ is the change of the energy inside the horizon. So from
equation (21) we have the amount of the energy crossing on the
horizon as \cite{Zhang}

\begin{equation}
-dE_{h}=-(n+1)\Omega_{n+1}r_{h}^{n}(\sigma +\dot{r}_{h}\rho)dv
\end{equation}

The first law of thermodynamics (Clausius relation) on the horizon
is defined as follows:

\begin{equation}
T_{h}dS_{h}=dQ=-dE_{h}
\end{equation}

From these equations, the time variation of entropy on the horizon
is given by

\begin{equation}
T_{h}\dot{S}_{h}=-(n+1)\Omega_{n+1}r_{h}^{n}(\sigma
+\dot{r}_{h}\rho)=-\frac{n(n+1)}{2}\Omega_{n+1}(\dot{m}+m'\dot{r}_{h})
\end{equation}

From above result, we may conclude that there is no role of
pressure of the perfect fluid on the variation of horizon entropy
for Vaidya space-time. Using (8), (24) and (27) we obtain the rate
of change of total entropy as

\begin{equation}
\dot{S}_{I}+\dot{S}_{h}=\frac{\Omega_{n+1}}{2T_{h}}\left[
n\dot{m}'r_{h}-n(n+1)\dot{m}- \{n^{2}m'+(n+1)m''r_{h}\}\dot{r}_{h}
\right]
\end{equation}

\subsection{\normalsize\bf{Apparent horizon}}

The dynamical apparent horizon $r_{A}$ can be found from
$h_{00}=0$ \cite{Cai3}, i.e.,

\begin{equation}
r_{A}^{n-1}=m(v,r_{A})
\end{equation}

The rate of change of total entropy (using (28) and (29)) for
apparent horizon is

\begin{equation}
\dot{S}_{I}+\dot{S}_{A}=\frac{\Omega_{n+1}}{2T_{A}}\left[
n\dot{m}'(v,r_{A})r_{A}-n(n+1)\dot{m}(v,r_{A})-\{n^{2}m'(v,r_{A})+(n+1)m''(v,r_{A})r_{A}\}\frac{\dot{m}(v,r_{A})}{(n-1)r_{A}^{n-2}}
\right]
\end{equation}

If the r.h.s. of the above expression is non-negative, then we can
say that the second law for the apparent horizon will be
fulfilled. For barotropic EOS, the mass function $m(v,r)$ is
defined in equation (10). In particular, we consider dark energy
dominated space-time (say, $k=-2/3$), the figure 1 shows the
variation of $\dot{S}_{I}+\dot{S}_{A}$ against $v$ for some
dimensions ($n=2$ (i.e., 4D), $n=3$ (i.e., 5D) and $n=4$ (i.e.,
6D)). The graphical representation of above expression shows that
the variation of total entropy for apparent horizon is positive
oriented. Thus the second law of thermodynamics for apparent
horizon is valid for all dimensions when we use the first law of
thermodynamics.\\

\begin{figure}
\includegraphics[scale=1]{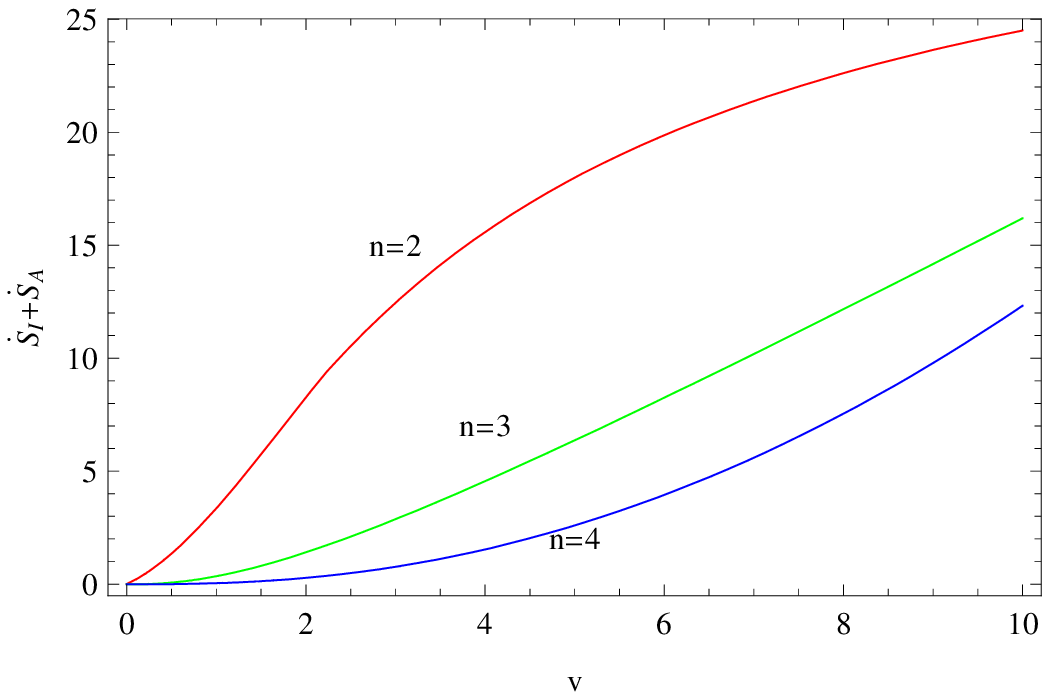}\\
\vspace{1mm} Fig.1\\

\vspace{6mm} Fig. 1 shows the variation of
$\dot{S}_{I}+\dot{S}_{A}$ against $v$ from eq.(30) for barotropic
fluid with $k=-2/3,f_{0}=1,g_{0}=1$. The red, green and blue lines
denote for $n=2$ (i.e., 4D), $n=3$ (i.e., 5D) and $n=4$ (i.e., 6D)
respectively.

 \vspace{6mm}
\end{figure}

\subsection{\normalsize\bf{Event horizon}}

The radius of the event horizon can be found from the metric (1),
i.e. \cite{Cai3}, ($ds^{2}=0=d\Omega_{n}^{2}$)

\begin{equation}
-\left(1-\frac{m(v,r)}{r^{n-1}}\right)dv^{2}+2dvdr=0
\end{equation}

or,
\begin{equation}
r_{E}^{n-1}=\frac{m(v,r_{E})}{1-2\dot{r}_{E}}
\end{equation}

The rate of change of total entropy (using (28) and (32)) for
event horizon can be obtained as

\begin{equation}
\dot{S}_{I}+\dot{S}_{E}=\frac{\Omega_{n+1}}{2T_{E}}\left[
n\dot{m}'(v,r_{E})r_{E}-n(n+1)\dot{m}(v,r_{E})-\frac{1}{2}\left(1-\frac{m(v,r_{E})}{r_{E}^{n-1}}
\right)\{n^{2}m'(v,r_{E})+(n+1)m''(v,r_{E})r_{E}\}\right]
\end{equation}

If the r.h.s. of the above expression is non-negative, then we can
say that the second law for the event horizon will be fulfilled.
Figure 2 shows the variation of $\dot{S}_{I}+\dot{S}_{E}$ against
$v$ for some dimensions ($n=2$ (i.e., 4D), $n=3$ (i.e., 5D) and
$n=4$ (i.e., 6D)) for dark energy dominated space time (say,
$k=-2/3$). The graphical representation of above expression shows
that the second law of thermodynamics for event horizon can not be
satisfied for all dimensions when we use the first law of thermodynamics.\\

\begin{figure}
\includegraphics[scale=1]{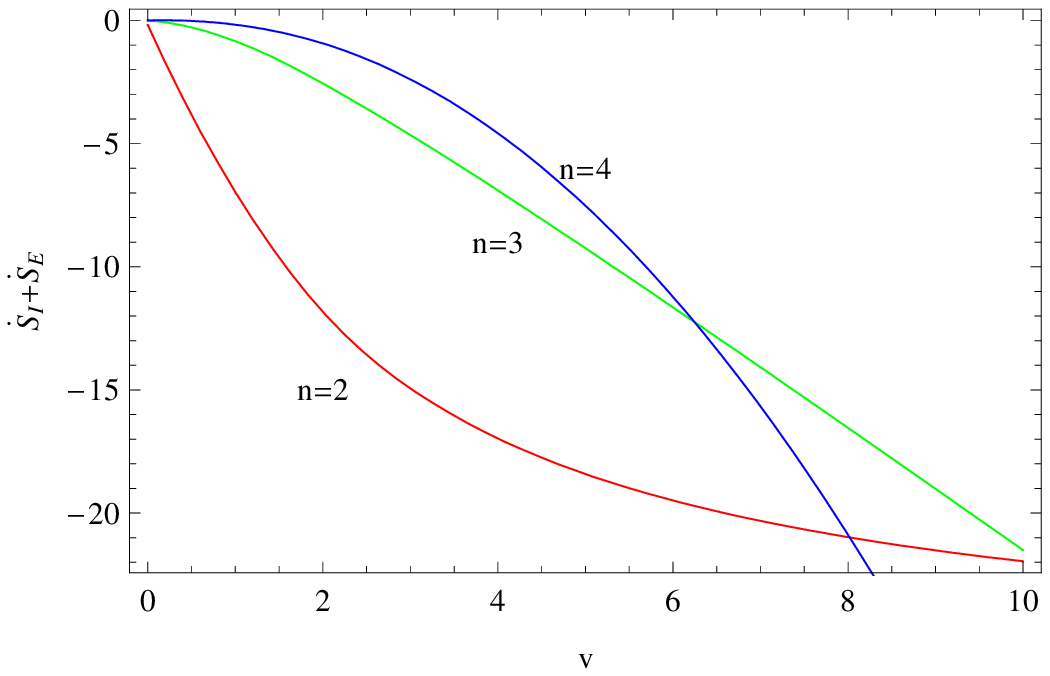}\\
\vspace{1mm} Fig.2\\

\vspace{6mm} Fig. 2 shows the variation of
$\dot{S}_{I}+\dot{S}_{E}$ against $v$ from eq.(33) for barotropic
fluid with $k=-2/3,f_{0}=1,g_{0}=1$. The red, green and blue lines
denote for $n=2$ (i.e., 4D), $n=3$ (i.e., 5D) and $n=4$ (i.e., 6D)
respectively.

 \vspace{6mm}
\end{figure}

\section{\normalsize\bf{GSL using area law}}

In this section we have not considered the first law of
thermodynamics, so we discard equation (27) for horizon entropy.
In this case, we shall consider the horizon entropy and horizon
temperature which are in the following.

\subsection{\normalsize\bf{Apparent horizon}}

The surface gravity is defined as \cite{Cai4, Zhou} follows

\begin{equation}
\kappa=\frac{1}{2\sqrt{-h}}\partial_{a}(\sqrt{-h}h^{ab}\partial_{b}(r))=\frac{(n-1)m-rm'}{2r^{n}}
\end{equation}

Hawking temperature on the apparent horizon is (using (34)) given
by

\begin{equation}
T_{A}=\frac{\kappa}{2\pi}=\frac{(n-1)m-rm'}{4\pi
r_{A}^{n}}=\frac{n-1}{4\pi r_{A}}-\frac{m'(v,r_{A})}{4\pi
m(v,r_{A})}
\end{equation}

For simple Vaidya space-time model (only null radiation),
$\rho=p=0$ i.e., $m'=0$, we get the surface gravity on the
apparent horizon as

\begin{equation}
\kappa=\frac{(n-1)m(v,r_{A})}{2r_{A}^{n}}=\frac{n-1}{2r_{A}}
\end{equation}

and in this case the temperature on the apparent horizon will be

\begin{equation}
T_{A}=\frac{n-1}{4\pi r_{A}}
\end{equation}

The entropy on the apparent horizon can be found from standard
Bekenstein-Hawking area law and is given by

\begin{equation}
S_{A}=\frac{{\cal A}}{4}=\frac{1}{4}(n+1)\Omega_{n+1} r_{A}^{n}
\end{equation}

Using (24), (29), (35) and (38), the rate of change of total
entropy for apparent horizon is obtained as

\begin{eqnarray*}
\dot{S}_{I}+\dot{S}_{A}=\frac{\Omega_{n+1}}{2T_{A}}\left[\frac{\{nm'(v,r_{A})-(n+1)m''(v,r_{A})r_{A}\}\dot{m}(v,r_{A})}{(n-1)r_{A}^{n-2}}
+n\dot{m}'(v,r_{A})r_{A} \right.
\end{eqnarray*}
\begin{equation}
\left. ~~~~~~~~~~
+\frac{n(n+1)}{8\pi}\dot{m}(v,r_{A})\left\{1-\frac{rm'(v,r_{A})}{(n-1)m(v,r_{A})}\right\}
\right]
\end{equation}

If the r.h.s. of the above expression is non-negative, then we can
say that the second law for the apparent horizon will be
fulfilled. Figure 3 shows the variation of
$\dot{S}_{I}+\dot{S}_{A}$ against $v$ for some dimensions ($n=2$
(i.e., 4D), $n=3$ (i.e., 5D) and $n=4$ (i.e., 6D)) for dark energy
dominated space time (say, $k=-2/3$). The graphical representation
of above equation shows that the second law of thermodynamics for
apparent horizon is satisfied for all dimensions.\\

\begin{figure}
\includegraphics[scale=1]{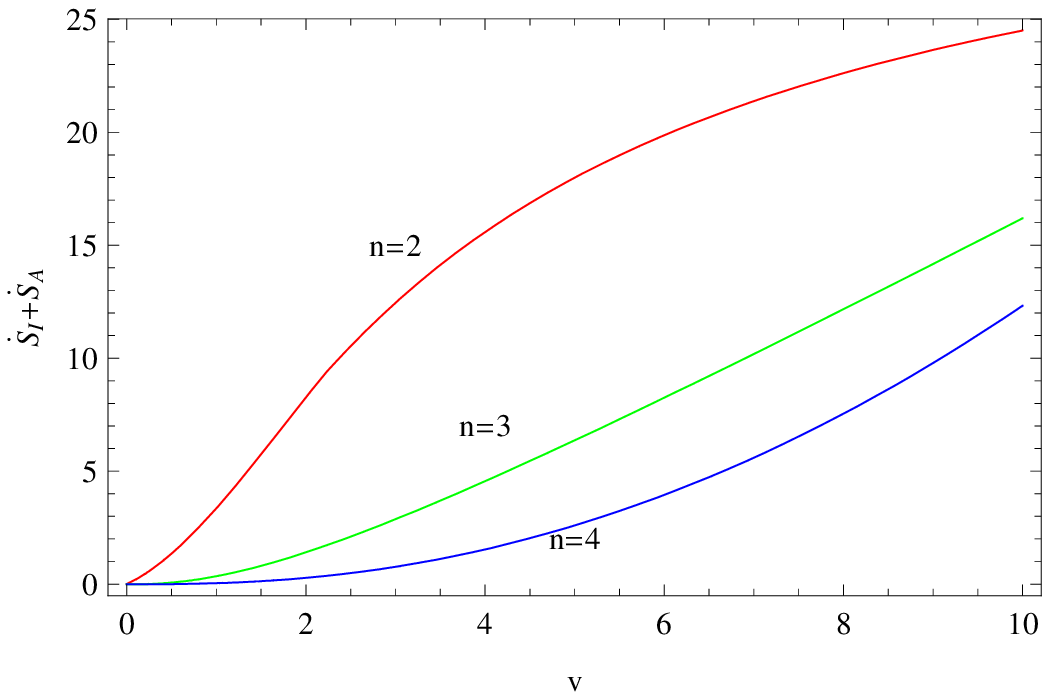}\\
\vspace{1mm} Fig.3\\

\vspace{6mm} Fig. 3 shows the variation of
$\dot{S}_{I}+\dot{S}_{A}$ against $v$ from eq.(39) for barotropic
fluid with $k=-2/3,f_{0}=1,g_{0}=1$. The red, green and blue lines
denote for $n=2$ (i.e., 4D), $n=3$ (i.e., 5D) and $n=4$ (i.e., 6D)
respectively.

 \vspace{6mm}
\end{figure}

\subsection{\normalsize\bf{Event horizon}}

The temperature on the event horizon is calculated as
\cite{Zhou,Zhao,Li,Vag}

\begin{equation}
T_{E}=\frac{(n-1)r_{E}^{n-2}(1-2\dot{r}_{E})-m'}{4\pi
m}=\frac{n-1}{4\pi r_{E}}-\frac{m'(v,r_{E})}{4\pi m(v,r_{E})}
\end{equation}

The entropy on the event horizon is

\begin{equation}
S_{E}=\frac{{\cal A}}{4}=\frac{1}{4}(n+1)\Omega_{n+1} r_{E}^{n}
\end{equation}

Using (24), (32), (40) and (41), the rate of change of total
entropy for event horizon is obtained as

\begin{eqnarray*}
\dot{S}_{I}+\dot{S}_{E}=\frac{\Omega_{n+1}}{2T_{E}}\left[\frac{1}{2}\left(1-\frac{m(v,r_{E})}{r_{E}^{n-1}}
\right)\{nm'(v,r_{E})-(n+1)m''(v,r_{E})r_{E}\}
+n\dot{m}'(v,r_{E})r_{E} \right.
\end{eqnarray*}
\begin{equation}
\left. +\frac{n(n+1)}{16\pi}\left(r_{E}^{n-1}-m(v,r_{E})
\right)\left(\frac{n-1}{r_{E}}-\frac{m'(v,r_{E})}{m(v,r_{E})}
\right)\right]
\end{equation}

If the r.h.s. of the above expression is non-negative, then we can
say that the second law for the event horizon will be fulfilled.
Figure 4 shows the variation of $\dot{S}_{I}+\dot{S}_{E}$ against
$v$ for some dimensions ($n=2$ (i.e., 4D), $n=3$ (i.e., 5D) and
$n=4$ (i.e., 6D)) for dark energy dominated space time (say,
$k=-2/3$). The graphical representation of above equation shows
that the second law of thermodynamics for event horizon can not be
satisfied for any dimensions when we use the area law of thermodynamics.\\

\begin{figure}
\includegraphics[scale=1]{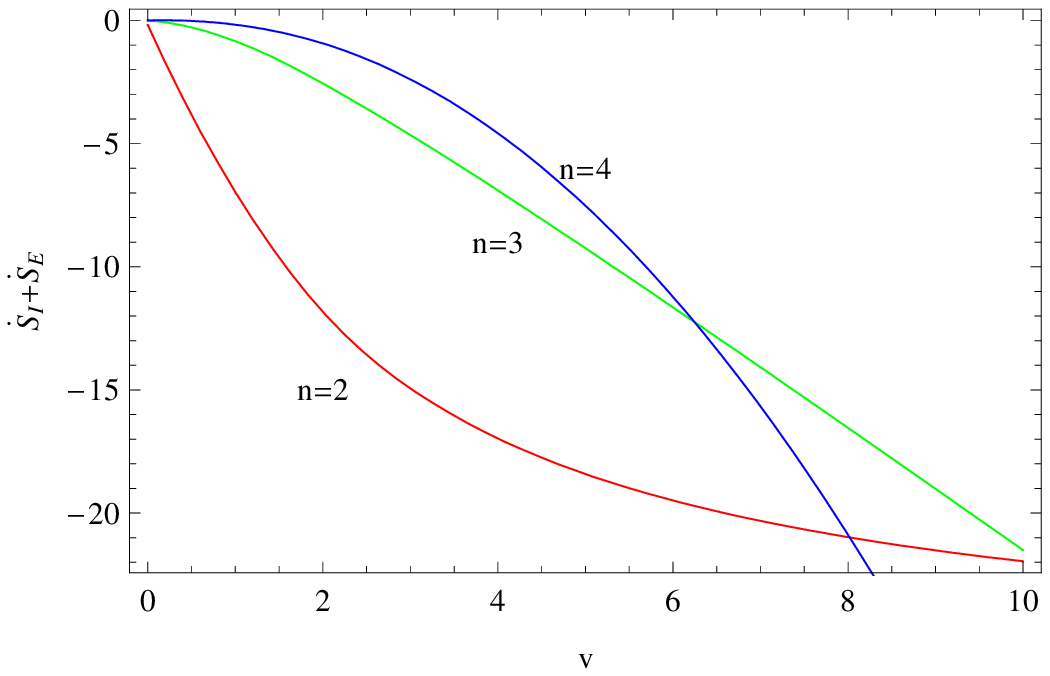}\\
\vspace{1mm} Fig.4\\

\vspace{6mm} Fig. 4 shows the variation of
$\dot{S}_{I}+\dot{S}_{E}$ against $v$ from eq.(42) for barotropic
fluid with $k=-2/3,f_{0}=1,g_{0}=1$. The red, green and blue lines
denote for $n=2$ (i.e., 4D), $n=3$ (i.e., 5D) and $n=4$ (i.e., 6D)
respectively.

 \vspace{6mm}
\end{figure}

\section{\normalsize\bf{Conclusions}}

In this work, we have considered the Vaidya spacetime in null
radiating fluid with perfect fluid in $(n+2)$-dimensions and have
found the solution for barotropic fluid with equation of state
$p=k\rho$. This generalized Vaidya solution is also known as
Husain solution, which is physically realistic for $k\ne 1/n$. The
solution contains two arbitrary functions $f(v)$ and $g(v)$, so
without any loss of generality, we have assumed (suitably)
$f(v)=f_{0}v^{n-1}$ and $g(v)=g_{0}v^{n(k+1)-2}$, where $f_{0}$
and $g_{0}$ are chosen to be positive. We have shown that the
Einstein's field equations can be obtained from Unified first law
i.e., field equations and unified first law are equivalent. The
first law of thermodynamics has also been constructed by Unified
first law. From this, the variation of entropy function has been
derived on the horizon. The variation of entropy function inside
the horizon has been derived using Gibb's law of thermodynamics.
So the total variation of entropy function has been constructed at
apparent and event horizons both. If we do not assume the first
law, then the entropy on the both horizons can be considered by
area law and the variation of total entropy has been found at both
the horizons. Also the validity of generalized second law (GSL) of
thermodynamics has been examined at both apparent and event
horizons by using the first law and using the area law separately.
When we use first law of thermodynamics, the GSL for apparent
horizon in any dimensions is satisfied, but the GSL for event
horizon can not be satisfied in any dimensions. For
Bekenstein-Hawking area law of thermodynamics, the GSL for
apparent horizon is satisfied in all dimensions the GSL for
event horizon can not be satisfied in any dimensions.\\\\

{\bf Acknowledgement:}\\

The author is thankful to IUCAA, Pune, India for providing
Associateship Programme under which part of the work was carried
out. The author also thanks to the members of Relativity and
Cosmology Research Centre, Jadavpur University, India for some
illuminating discussions.\\

\end{document}